\documentclass[preprint, preprintnumbers,amsmath,english,amssymb]{revtex4}
\usepackage[T1]{fontenc}
\usepackage[latin9]{inputenc}
\usepackage{color}
\usepackage{amsmath}
\usepackage{graphicx}
\usepackage{amssymb}

\makeatletter

\@ifundefined{textcolor}{}
{%
 \definecolor{BLACK}{gray}{0}
 \definecolor{WHITE}{gray}{1}
 \definecolor{RED}{rgb}{1,0,0}
 \definecolor{GREEN}{rgb}{0,1,0}
 \definecolor{BLUE}{rgb}{0,0,1}
 \definecolor{CYAN}{cmyk}{1,0,0,0}
 \definecolor{MAGENTA}{cmyk}{0,1,0,0}
 \definecolor{YELLOW}{cmyk}{0,0,1,0}
 }

\makeatother

\usepackage{babel}

\begin{document}

\title{Energy-gap Opening and Quenching in Graphene under Periodic External
Potentials}

\author{Aihua Zhang$^{1}$}

\author{Zhenxiang Dai$^{1}$}

\author{Lei Shi$^{1,3}$}

\author{Yuan Ping Feng$^{1}$}

\author{Chun Zhang$^{1,2}$}

\email{phyzc@nus.edu.sg}

\affiliation{$^{1}$Department of Physics,National University of Singapore, 2
Science Drive 3, Singapore,117542\\
 $^{2}$Department of Chemistry, National University of Singapore,
3 Science Drive 3, Singapore, 117543\\
 $^{3}$Department of Physics and Surface Physica Laboratory, Fudan
University, Shanghai, China, 200433}
\date{\today}

\begin{abstract}
In this letter, we investigated the effects of periodic external potentials on properties of charge carriers in graphene using both the first-principles method based on density functional theory (DFT) and a theoretical approach based on a generalized effective spinor Hamiltonian. 
DFT calculations were done in a modified Kohn-Sham procedure that includes effects of the
periodic external potential. Unexpected energy band gap
opening and quenching were predicted for the graphene superlattice with two symmetrical sublattices and those with two unsymmetrical sublattices, respectively. Theoretical analysis based on the spinor Hamiltonian showed that the correlations between pseudospins of Dirac fermions in graphene and the applied external potential, and the potential-induced intervalley scattering, play important roles in energy-gap opening and quenching.
\end{abstract}

\maketitle

Graphene has attracted great interests due to its peculiar electronic properties~\cite{NovoselovSCI, NovoselovNAT, KimNAT, KimPRL, DeheerSCI, Duanxf, NetoREVIEW}. Recently there has been a surge in research pertaining to
properties of charge carriers in graphene under periodic external
potentials. Based on the commonly used two-component effective Spinor Hamiltonian~\cite{ParkNP}, periodic external potentials
were found to have significant effects on behaviors of low-energy
quasi-particles in graphene (i.e.\ the massless Dirac fermions),
leading to exciting new phenomena such as anisotropic group velocities
of Dirac fermions~\cite{ParkNP}, emerging zero-energy states~\cite{Fertig09},
new massless charge carriers~\cite{ParkPRL},
unusual Landau levels and quantum Hall effects~\cite{ParkQHE},
and the supercollimation of electron beam in graphene superlattice~\cite{ParkNL},
all of which suggested a promising direction for future design
of graphene-based electronic devices without the need for cutting
or etching. 

The linear energy dispersion of charge carriers in graphene near the
Dirac points enables us to describe the behaviors of low-energy quasi-particles
in the vicinity of each Dirac point using a simple two-component effective
spinor Hamitonian~\cite{Divincenzo,Ando}, \begin{equation}
\mathcal{H}_{0}(\mathbf{k})=\hbar v_{0}\left(\begin{array}{cc}
0 & \hat{k}_{x}-i\hat{k}_{y}\\
\hat{k}_{x}+i\hat{k}_{y} & 0\end{array}\right),\label{eq:ehsv}\end{equation}
where $v_{0}$ is the Fermi velocity. The eigenstates of this spinor
Hamiltonian have a degree of freedom of pseudospin, $s\in\{\uparrow,\downarrow\}$,
originating from two sublattices A and B in graphene, and can be expressed
as $\mid s,\mathbf{k}+\mathbf{K}\rangle$, where $\mathbf{K}$ is
the reference Dirac point. When the graphene is subject to a periodic
external potential, the aforementioned commonly used effective Hamitonian~\cite{ParkNP,ParkQHE,Fertig09,ParkPRL,ParkNL}
is 
\begin{equation}
\mathcal{H}(\mathbf{k})=\mathcal{H}_{0}(\mathbf{k})+U(r)\mathcal{I},\label{eq:ehplain}
\end{equation}
where $U(r)$ is the external potential, and $\mathcal{I}$ is a $2\times2$
unit matrix. This effective Hamiltonian has been widely used in various types of graphene-based
supperlattice structures~\cite{ParkNP, Fertig09, ParkPRL, ParkQHE, ParkNL, Pedersen}.
This Hamiltonian neglects the charge redistribution in
graphene induced by the external potential, the interaction between the external potential and two different pseudospin channels, and the intervalley
scattering between different Dirac points, all of which are assumed to be unimportant when the potential size and periodicity
are big ($\sim10\mbox{ nm}$). As the current technology
makes it possible to fabricate electric gates of intermediate size
in a few nanometers~\cite{Chou}, it is therefore important to investigate
all these neglected effects on properties of charge carriers in graphene
superlattice, and to test their convergence for potential periodicity close to 10 nm.

\begin{figure}
\includegraphics{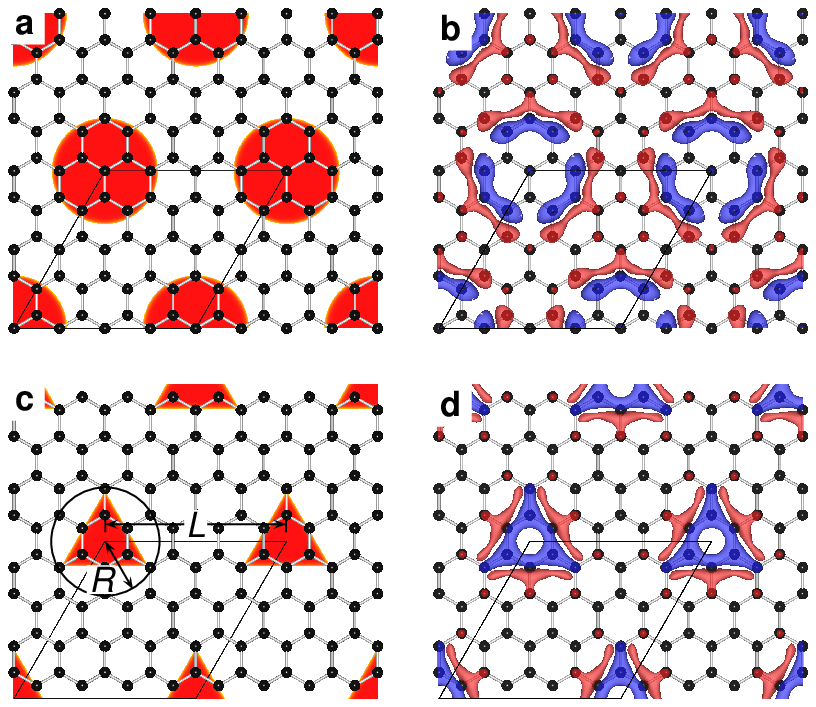}

\caption{\label{fig:model} Graphene superlattices and charge redistribution.
\textbf{a}, 2D (4$\times$4) graphene superlattice with a circular
muffin-tin type of potential on the top position. \textbf{b}, The
DFT charge redistribution, $\rho^{V}(\mathbf{r})-\rho^{0}(\mathbf{r})$,
due to the potential in \textbf{a}. The red (blue) isosurface corresponds
to a charge density difference of $4\times10^{-3}$~$e$/\AA$^3$
($-4\times10^{-3}$~$e$/\AA$^3$). \textbf{c}, Same as \textbf{a}
but for a triangular potential and the hollow position. \textbf{d}, Similar
quantity as in \textbf{b} for the potential in \textbf{c}.}

\end{figure}

In current work, two different shapes of periodic external potentials, circle and triangle
(as shown in Fig.~\ref{fig:model}), were considered.
Each of these applied external potentials has a 2-dimentional (2-d) $(N\times N)$ triangular
periodicity. For each potential shape, we considered two possible
center positions: on top or on hollow. In Fig.~1,
we show two examples, a circular potential centered on top (Fig.~1a)
and a triangular potential centered on hollow (Fig.~1c),
each of which has a $(4\times4)$ superlattice periodicity. The potential
size is denoted as $R$, and the superlattice periodicity $L$ equals
to $N\times a_{0}$, where $a_{0}$ is the lattice constant of pristine
graphene. In one supercell, the external potential is assumed to have
the form $V_{ext}(x,y)=V_{0}/(e^{d/\Delta}+1)$, where $d$ is the
nearest distance between the considered point in graphene plane $(x,y)$
and the pre-defined potential edges ($d$ is negative if the considered point is inside the
potential). Throughout the paper, the smearing distance $\Delta$
is taken to be 0.1~\AA{}. When crossing the edges of the potential
from inside to outside, $V_{ext}$ continuously and rapidly changes
from $V_{0}$ to zero. Along the direction normal to the graphene
plane, the external potential is assumed invariant. It is worthy noting
here that for above-mentioned circular and triangular potentials,
the widely used effective Hamiltonian described in Eq.~\ref{eq:ehplain}
gives zero energy band gaps regardless of center positions, the potential
size $R$, and the periodicity $L$. 

For small systems ($L<3nm$), first-principles calculations based on DFT are possible. In order to take into account the effects of external potential, we modified the Vienna \textit{ab initio} simulation package~\cite{vasp1,vasp2} to read in aforementioned external potential $V_{ext}$ and add the potential to the system's Kohn-Sham Hamiltonian. In Fig.~1b and Fig.~1d, we showed the external-potential-induced charge redistribution from DFT calculations in two graphene superlattice structures as shown in Fig.~1a and Fig.~1c, respectively. We will see later in this paper that the potential-induced charge redistribution has significant effects on properties of charge carriers in graphene. Details of DFT calculations can be found in reference~\cite{DFT}.

First, we
calculated electronic properties of various small graphene superlattice
structures with a fixed $(4\times4)$ periodicity for both circular and triangular external potentials using
aforementioned DFT method. In all these calculations, the potential
strength $V_{0}$ is fixed to 1.0 eV. The energy band gaps from DFT
as functions of potential size $R$ for different superlattice structures
are shown in Fig.~2. Except for the circular potential
centered on hollow position (\texttt{CH} in the figure) which has
a zero energy gap regardless of the potential size, all superlattice
structures present non-negligible energy gaps for most values of potential
size, and show complicated behaviors of energy gap when the potential
size varies, which can not be understood by the aforementioned widely used effective Hamiltonian.

We then generalized the widely used spinor Hamiltonian to take into account previously-neglected effects such as the charge redistribution, the pseudospin-potential correlation, and the potential-induced intervalley scattering. Based on the generalized spinor Hamiltonian, we showed that both the PseudoSpin-potential
correlation (PS) and the Intervalley Scattering (IS) have great effects on
behaviors of Dirac fermions, resulting in unexpected opening or quenching
of energy band gap in the graphene superlattice.

In order to take into account the external-potential induced intervalley
scattering, we consider the following $4\times4$ spinor Hamiltonian
for pristine graphene~\cite{Ando}, which is essentially the combination
of two independent single-valley Hamiltonian (see Eq.~\ref{eq:ehsv})
for two inequivalent Dirac points, $\mathbf{K}$ and $\mathbf{K'}$,
\begin{equation}
\mathcal{H}_{0}(\mathbf{k})=\hbar v_{0}\left(\begin{array}{cccc}
0 & -\omega(\hat{k}_{x}-i\hat{k}_{y}) & 0 & 0\\
-\omega^{\dagger}(\hat{k}_{x}+i\hat{k}_{y}) & 0 & 0 & 0\\
0 & 0 & 0 & \hat{k}_{x}+i\hat{k}_{y}\\
0 & 0 & \hat{k}_{x}-i\hat{k}_{y} & 0\end{array}\right),\label{eq:eh4x4}\end{equation}
where $\omega=\mathrm{e}^{-2\pi i/3}$. Then, the potential-induced
scattering between different pseudospin channels and different Dirac
points can be evaluated as, \[
\mathcal{U}_{s',\mathbb{K}';s,\mathbb{K}}=\langle s',\mathbb{K}'\mid U(\mathbf{r})\mid s,\mathbb{K}\rangle,\] where \[ s\in\{\uparrow,\downarrow\},\:\mathbb{K}\in\{\mathbf{K},\,\mathbf{K}'\}.\]
Under the framework of the single-orbital ($2p_{z}$ of carbon) tight-binding
approach~\cite{NetoREVIEW}, the potential-induced scattering can be worked
out after some simple algebra, \begin{equation}
\mathcal{U}_{s',\mathbb{K}';s,\mathbb{K}}=\delta_{s',s}\sum_{\mathbf{R}_{s}}\mathrm{e}^{-i(\mathbb{K}'-\mathbb{K})\cdot\mathbf{R}_{s}}\bar{\rho}_{2p_{z}}(\mathbf{r}-\mathbf{R}_{s})\cdot U(\mathbf{r}).\label{eq:umatrix}\end{equation}
In above equation, the summation is over the sublattice A $(\mathbf{R}_{\uparrow})$
or B $(\mathbf{R}_{\downarrow})$ of graphene. $\bar{\rho}_{2p_{z}}(\mathbf{r})$
is the charge density of carbon $2p_{z}$ orbital averaged along $z$
direction (perpendicular to graphene), which is approximated by
a 2-dimensional Gaussian function $\frac{1}{2\pi\sigma^{2}}\mathrm{e}^{-\frac{|\mathbf{r}|^{2}}{2\sigma^{2}}}$
with $\sigma=0.25\mbox{ \AA}$ in this work. $\delta_{s',s}$ is present because
of the neglection of pseudospin flip. Combining \ref{eq:eh4x4}
and \ref{eq:umatrix}, we obtain the effective Hamiltonian for graphene
under external potential with the pseudospin-potential correlation
 and the intervalley scattering, \begin{equation}
\mathcal{H}_{PS-IS}(\mathbf{k})=\mathcal{H}_{0}(\mathbf{k})+c_{scf}\mathcal{U},\label{eq:ehpsis}\end{equation}
where the constant $c_{scf}$ is introduced to account for effects
of self-consistent field such as potential-induced charge redistribution
which is not included in the derivation of the scattering matrix.
This parameter can be determined by fitting DFT results for small
systems. If the fitted $c_{scf}$ is close to one, the effects of self-consistent
field are not important. Otherwise,
it can not be neglected.

For a periodic external potential which has a two dimensional $(N\times N)$ periodicity with respect to the primitive
cell of graphene,
the intervalley scattering terms in Eq.~\ref{eq:umatrix}
(those for $\mathbb{K}'\neq\mathbb{K}$) are only significant when
$N$ is multiples of $3$ due to the phase $\mathrm{e}^{-i(\mathbb{K}'-\mathbb{K})\cdot\mathbf{R}_{s}}$.
Therefore, for $N$ is not multiples of $3$, the effective Hamiltonian
$\mathcal{H}_{PS-IS}$ reduces to
\begin{equation}
\mathcal{H}_{PS}(\mathbf{k})=\hbar v_{0}\left(\begin{array}{cc}
0 & \hat{k}_{x}-i\hat{k}_{y}\\
\hat{k}_{x}+i\hat{k}_{y} & 0\end{array}\right)+c_{scf}\left(\begin{array}{cc}
U_{\uparrow}(\mathbf{r}) & 0\\
0 & U_{\downarrow}(\mathbf{r})\end{array}\right),\label{eq:ehps}
\end{equation}
where $\mathbf{U}_{s\in\{\uparrow,\downarrow\}}(\mathbf{r})=\sum_{\mathbf{R}_{s}}\bar{\rho}_{2p_{z}}(\mathbf{r}-\mathbf{R}_{s})\cdot U(\mathbf{r})$.
In the rest of the paper, effective Hamiltonians $\mathcal{H}_{PS-IS}$ and $\mathcal{H}_{PS}$
are represented by EH-PS-IS and EH-PS respectively. These two Hamiltonians are suitable
for potentials of arbitrary sizes. They can be diagonalized in the same way
as diagonalizing the commonly used Hamiltonian (Eq.~\ref{eq:ehplain})
by expanding the Hamiltonian in plane-wave basis~\cite{ParkNP,ParkQHE,Fertig09,ParkPRL,ParkNL}.

For small superlattice structures with $(4 \times 4)$ periodicity, we applied the effective Hamiltonian with the pseudospin-potential
correlation (EH-PS) as described in Eq.~\ref{eq:ehps}, and found good agreement with
DFT results for all systems when setting the parameter $c_{scf}$
to 1.6 as shown in Fig. 2. The zero energy gap for the \texttt{CH} potential can be easily
understood by the fact that in this case, the external potential interacts
with two different pseudospin channels in the exactly same way, resulting
in the identical $U_{\uparrow}$ and $U_{\downarrow}$ in $\mathcal{H}_{PS}$.
Therefore in this case, the EH-PS is essentially the same as the conventionally
used effective Hamiltonian as described in Eq.~\ref{eq:ehplain},
which gives zero energy gap for all cases. For other potentials (\texttt{CT},
\texttt{TT} and \texttt{TH} in the figure), the symmetry between the
up and down pseudospins is broken, leading to different $U_{\uparrow}$
and $U_{\downarrow}$ in $\mathcal{H}_{PS}$, and the energy-gap opening.
The value of $c_{scf}$, 1.6, clearly shows the importance of the
effects of the self-consistent field. Actually, if neglecting these
effects by setting $c_{scf}=1$, the energy gaps are generally underestimated
by EH-PS by about 50\%.

We then change the potential periodicity $L$ from 1~nm to 8~nm
to examine the $L$-dependence of the energy band gap. In these calculations,
$V_{0}$ is set to 1~eV as before, and the potential size is taken
to be $R=\dfrac{3}{4a_{0}}L$ for triangular potentials, and $R=\dfrac{1}{4}L$
for circular potentials. DFT calculations were performed for $L<3\mbox{ nm}$.
Results are presented in Fig.~3. For triangular
potentials, regardless of center positions, the intervalley scattering
has negligible effects. For small systems, the EH-PS gives good results
compared to DFT, and for cases that the periodicity $N$ is multiples
of 3, where EH-PS-IS is applicable, two Hamiltonians, EH-PS and EH-PS-IS,
agree with each other quite well, indicating the fact that the intervalley
scattering for triangular potentials are not important regardless
of the system size. For circular potentials, the intervalley scattering
show great effects on energy gaps, and interestingly, these effects
are completely different when the center positions of potentials
are different. For the case of circular potential centered on top (\texttt{CT}),
the EH-PS tends to open significant energy gaps. When the intervalley
scattering is not turned on ($N$ is not the multiples of 3), EH-PS
gives good results compared to DFT for small systems. While, when
$N$ is multiples of 3, the inclusion of the intervalley scattering
(EH-PS-IS) greatly decreases or quenches the energy gap opened by
EH-PS. For small systems, $N=6,\,9,\,12$, the quenching
of the energy gap due to the intervalley scattering is confirmed by DFT calculations.
For the case of circular potential on hollow (\texttt{CH}), EH-PS
predicts zero energy gap regardless of the potential size and periodicity
due to the reserved symmetry between up and down pseudospins as mentioned before.
While, in this case, the presence of the intervalley scattering ($N$
is multiples of 3) opens significant energy gaps as shown in the figure.
The gap-opening by the intervalley scattering was verified by DFT
calculations for small systems when $N=6,\,9,\,12$. 
The energy-gap opening and quenching, as well as the
convergence of the energy-gap for large potential periodicity we presented
here clearly suggest that in order to correctly understand the behaviors
of quasi-particles in graphene under periodic external potential, the pseudospin-potential correlation (PS) and the intervalley scattering (IS) have to be properly included in the effective Hamiltonian.

The predicted energy-gap opening and quenching in graphene superlattice
are further illustrated in schematic diagrams shown in Fig.~4.
In Fig.~4a, we demonstrate the gap opening due to
the breaking of the symmetry between two pseudospin channels caused by
the pseudospin-potential correlation in EH-PS. In Fig.~4b,
an example of the energy-gap quenching due to the intervalley scattering
is shown. First, the pseudospin-potential correlation breaks the symmetry
between up and down pseudospins for both Dirac points $\mathbf{K}$
and $\mathbf{K}'$, and then the intervalley scattering breaks the
symmetry between $\mathbf{K}$ and $\mathbf{K}'$ for the same pseudospin,
resulting in two degenerate states in the middle, one for spin up
and one for spin down. This is exactly what happens in graphene under
\texttt{CT} potential when the potential periodicity $N$ is multiple
of 3. In Fig.~\ref{fig:physpic}c, we show the mechanism of the gap
opening due to the intervalley scattering for the case of \texttt{CH}
potential. In this case, the symmetry between two pseudospin channels
are reserved in EH-PS (i.e.\ in Eq.~\ref{eq:ehps}, $U_{\uparrow}$
is identical to $U_{\downarrow}$), and the intervalley scattering
breaks the symmetry between two Dirac points, leading to the gap opening.

In summary, in this paper, we investigated properties of Dirac Fermions in graphene under periodic external potential via both the first-principles method and a theoretical approach based on a generalized effective spinor Hamiltonian.
The generalized effective spinor
Hamiltonian takes into account the pseudospin-potential correlation
and the intervalley scattering, and is suitable for systems with arbitrary
potential size and periodicity. The intervalley scattering is found to be significant only when the potential periodicity $N$ is multiples of 3. 
Unexpected energy-gap opening and
quenching in graphene superlattice due to the interplay between the pseudospin-potential
correlation and the intervalley scattering are predicted. For small
systems, results from the generalized effective Hamitonian agree very
well with DFT calculations, and for large systems, the proposed Hamiltonian
gives qualitatively different results from the commonly used Hamiltonian
in previous studies. The generalization of the effective Hamiltonian to other graphene superlattice structures with complicated boundary conditions such as antidot lattice and graphene nanoribbon-based superlattice will be done in our future work.


This work was supported by NUS Academic Research Fund (Grant Nos:
R-144-000-237-133 and R-144-000-255-112). Computations were performed
at the Centre for Computational Science and Engineering at NUS.



\newpage

\newpage

\begin{figure}
\includegraphics{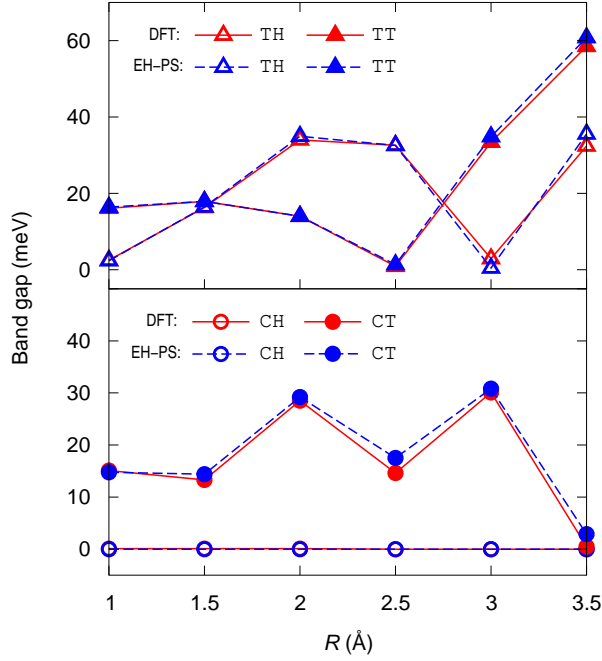}

\caption{\label{fig:gap_vs_r} Variation of band gaps as a function
of $R$. Band gaps were calculated using DFT and effective Hamiltonian
with pseudospin-potential correlation (EH-PS) with $V_{0}=1\mbox{ eV}$
and $L=4a_{0}$. The first character of legends refer to potential
shapes, with \texttt{C} and \texttt{T} standing for circle and triangle,
respectively, and the second character, \texttt{H} or \texttt{T},
specifies the hollow or top position for the potential center.}

\end{figure}

\newpage

\begin{figure}
\includegraphics{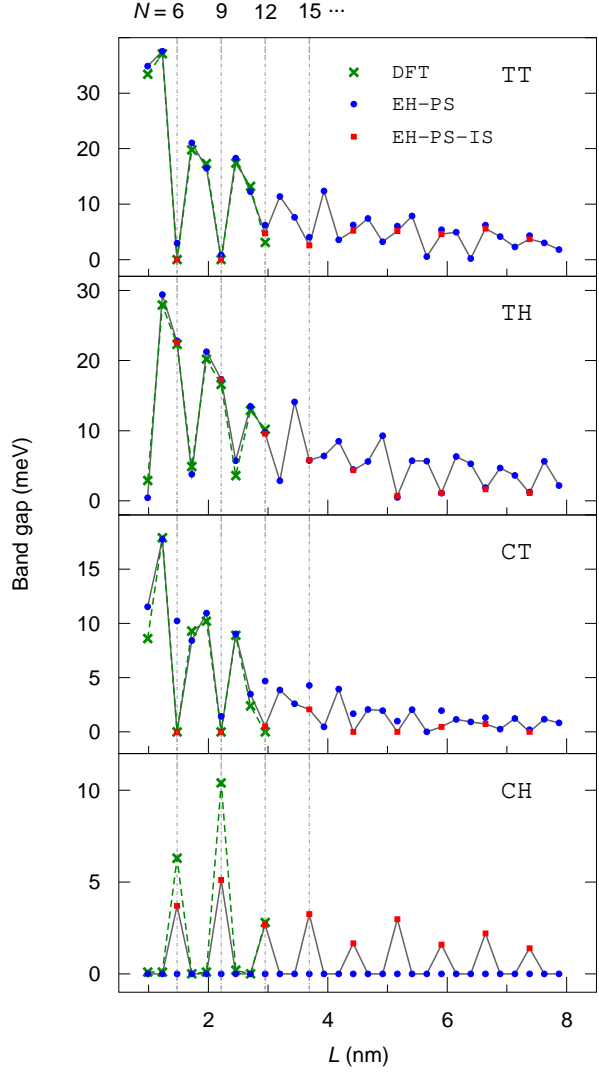}

\caption{\label{fig:gap_vs_l} Variation of band gaps as a function
of $L\:(=Na_{0})$. Band gaps were obtained from DFT, effective Hamiltonian
with pseudospin-potential correlation (EH-PS) and effective Hamiltonian
with pseudospin-potential correlation and inter-valley scattering
(EH-PS-IS). The denotation for different potentials is the same as
that in Fig~\ref{fig:gap_vs_r}. Other parameters are $V_{0}=1\mbox{ eV}$,
$R=\dfrac{3}{4a_{0}}L$ for triangular potentials and $R=\dfrac{1}{4}L$
for circular potentials.}

\end{figure}

\begin{figure}
\includegraphics{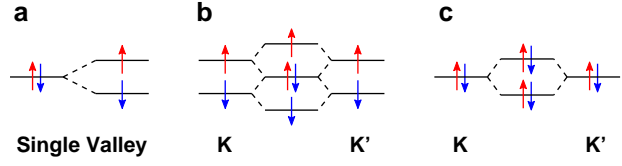}

\caption{\label{fig:physpic} Schematic diagrams of band-gap opening
and quenching in graphene superlattice. \textbf{a}, Due to the pseudospin-potential
correlation, the symmetry between up and down pseudospin channels is broken, and a gap is open.
\textbf{b}, The pseudospin-potential correlation breaks pseudospin symmetry at both Dirac points, $K$ and $K'$, and the intervalley scattering mixes different states with the same spin at two Dirac points, leading to two degenerate states in the middle and the quenching of the energy-gap. \textbf{c}, For the case that the symmetry between two pseudospin channels is reserved in the pseudospin-potential
correlation (CH potential), the mixing of two states with the same pseudospin at $K$ and $K'$ due to the intervalley scattering results in a energy gap.}

\end{figure}

\end{document}